\documentclass[conference]{IEEEtran}
\IEEEoverridecommandlockouts

\usepackage{cite}
\usepackage{amsmath,amssymb,amsfonts}
\usepackage{color,soul}
\usepackage{graphicx}
\usepackage{physics}
\usepackage{textcomp}
\usepackage{xcolor}
\usepackage{algorithm}
\usepackage{algpseudocode}
\usepackage{xspace}
\usepackage{hyperref}
\usepackage{multirow}
\usepackage{pifont}
\usepackage{tikz}
\usepackage{bm}
\usepackage[table,xcdraw]{}
\newcommand*\circled[1]{%
  \tikz[baseline=(char.base)]{
    \node[shape=circle,fill,inner sep=0pt,minimum size=1em] (char) 
    {\raisebox{0.15ex}{\vphantom{A}\hspace{0.1em}\textcolor{white}{#1}\hspace{0.1em}}};}}
    
\def\BibTeX{{\rm B\kern-.05em{\sc i\kern-.025em b}\kern-.08em
    T\kern-.1667em\lower.7ex\hbox{E}\kern-.125emX}}

\begin{document}

\title{Capturing Quantum Snapshots from a Single Copy via Mid-Circuit Measurement and Dynamic Circuit}

\author{\IEEEauthorblockN{Debarshi Kundu}
\IEEEauthorblockA{\textit{Dept of CSE} \\
\textit{Penn State University}\\
State College, USA \\
dqk5620@psu.edu}
\and
\IEEEauthorblockN{Avimita Chatterjee}
\IEEEauthorblockA{\textit{Dept of CSE} \\
\textit{Penn State University}\\
State College, USA\\
amc8313@psu.edu}
\and
\IEEEauthorblockN{Archisman Ghosh}
\IEEEauthorblockA{\textit{Dept of CSE} \\
\textit{Penn State University}\\
State College, USA\\
amc8313@psu.edu}
\and
\IEEEauthorblockN{Swaroop Ghosh}
\IEEEauthorblockA{\textit{School of EECS} \\
\textit{Penn State University}\\
State College, USA \\
szg212@psu.edu}
\and

}

\maketitle

\begin{abstract}
We propose Quantum Snapshot with Dynamic Circuit (QSDC), a hardware-agnostic, learning-driven framework for capturing quantum snapshots: non-destructive estimates of quantum states at arbitrary points within a quantum circuit, which can then be classically stored and later reconstructed. This functionality is vital for introspection, debugging, and memory in quantum systems, yet remains fundamentally constrained by the no-cloning theorem and the destructive nature of measurement.
QSDC introduces a guess-and-check methodology in which a classical model — powered by either gradient-based neural networks or gradient-free evolutionary strategies — is trained to reconstruct an unknown quantum state using fidelity from the SWAP test as the sole feedback signal. Our approach supports single-copy, mid-circuit state reconstruction, assuming hardware with dynamic circuit support and sufficient coherence time.
We validate core components of QSDC both in simulation and on IBM quantum hardware. In noiseless settings, our models achieve average fidelity up to 0.999 across 100 random quantum states; on real devices, we accurately reconstruct known single-qubit states (e.g., Hadamard) within three optimization steps.

\end{abstract}

\begin{IEEEkeywords}
Quantum Snapshot, Quantum Memory, Quantum State Tomography, Mid-circuit Measurement, Dynamic Circuit 
\end{IEEEkeywords}

\section{Introduction}

\begin{figure}[!hbp]
    \centering
    \includegraphics[width=0.8\linewidth]{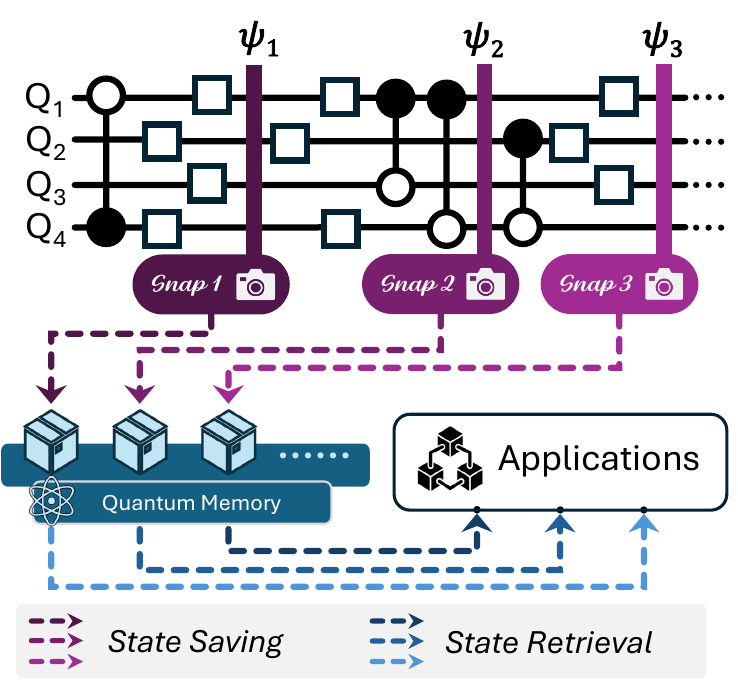}
    \caption{\textbf{Overview of Quantum Snapshot:} Quantum states are observed (snapshots) at various points within a single circuit at multiple intersections, without collapsing the original system. This enables non-destructive access to quantum information during circuit execution.
}
    \label{fig:quantum_bank}
\end{figure}

\begin{figure}
    \centering
    \includegraphics[width=1\linewidth]{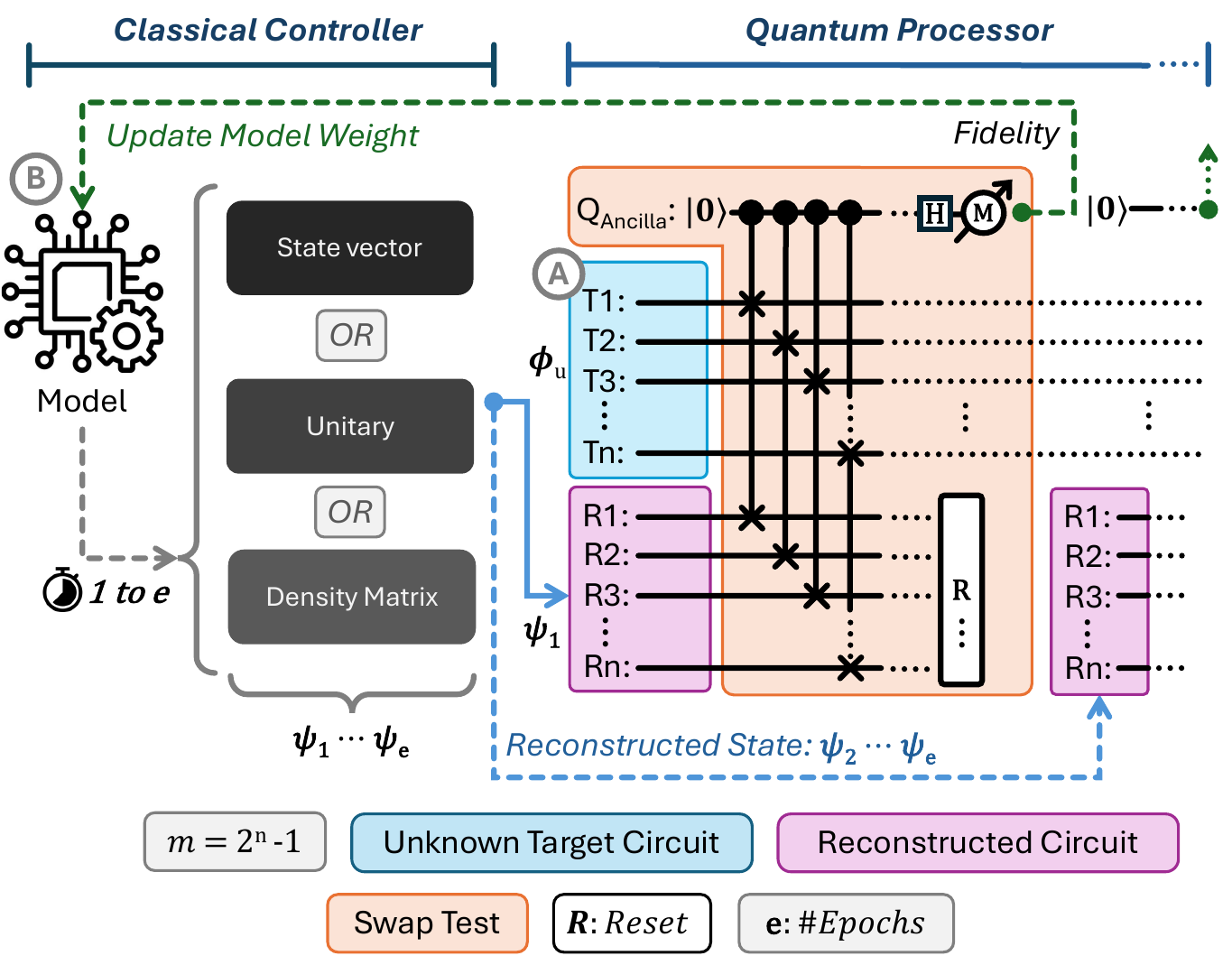}
    \caption{
    \textbf{Quantum Snapshot Framework with Mid-Circuit Measurement and Dynamic Circuit Execution.}
This schematic illustrates the QSDC pipeline for non-destructive quantum state reconstruction. An unknown quantum state (blue block) is compared against a classically generated guess (purple block) using a SWAP test mediated by an ancilla qubit. Mid-circuit measurements capture fidelity, which is fed back to update the classical model. Dynamic circuit features—such as qubit reset and reinitialization—enable iterative optimization within a single execution, allowing single-copy state estimation without collapsing the quantum state
}
    \label{fig:process}
\end{figure}

Identifying a quantum state mid-circuit without destroying it---often referred to as capturing a \textit{quantum snapshot}---remains beyond the capabilities of current quantum technology. Similarly, no reliable approach exists for indefinitely storing arbitrary quantum states. Yet, these functionalities are pivotal for advancing quantum technologies, with promising applications in quantum communication~\cite{heshami2016quantum}, control~\cite{riste2012feedback}, debugging~\cite{erhard2019characterizing}, and real-time circuit introspection~\cite{hebenstreit2015real}.

\subsection{Motivation}
At the heart of quantum computation lies a fundamental challenge: the ability to observe and understand the state of a quantum system. Unlike classical systems, where observation is passive and non-disruptive, quantum systems exhibit a profound fragility. The act of measurement does not merely reveal information, it irreversibly alters the very state being observed. This phenomenon, known as wavefunction collapse, ensures that once a quantum state is measured, its original form ceases to exist. As a result, direct measurements both destroy the intricate superpositions and entanglements that underpin quantum advantage and yield only partial information about the system.
A close neighbor of the quantum snapshot concept is Quantum State Tomography (QST). Traditionally, QST has been the principal technique for reconstructing a complete description of a quantum state, typically as a density matrix, by performing repeated measurements on many identically prepared copies of the system. It plays a critical role in verifying prepared states, benchmarking quantum hardware, and characterizing noise in quantum processes, and has become a standard tool for experimental validation. However, QST remains fundamentally destructive: it relies on projective measurements that collapse the quantum state during the process of reconstruction. Consequently, it cannot serve as a mechanism for capturing quantum snapshots, where the goal is to capture the state non-destructively and without requiring multiple copies. Besides, QST is a painstakingly slow and tedious process. Full QST typically requires \( 3^n \) distinct measurement settings—one for each combination of Pauli basis on \( n \) qubits—which implies the need for that many copies of the same quantum state~\cite{smith_efficient_2021}. As such, QST is unsuitable for real-time introspection or applications requiring non-destructive mid-circuit state access.
Equally pressing is the issue of \textit{quantum memory}. Today’s quantum technologies face severe coherence limitations, with qubits quickly losing their stored quantum information due to environmental noise~\cite{schlosshauer2019quantum, gundougan2023ultimate}. Although long-term quantum memory is a critical component for applications like quantum repeaters, distributed quantum computing, and the quantum internet~\cite{kimble2008quantum, sangouard2011quantum, wehner2018quantum}, it remains largely out of reach with current hardware.
Given these challenges, a natural and pressing question arises:  
\textbf{Can we extract quantum information at multiple intersections of the quantum circuit from a single copy \textit{without} collapsing the state, and store it classically for later retrieval during quantum computation?}
This work takes a step toward answering that question by proposing a new framework for non-destructive, single-copy quantum state estimation and memory.
\begin{figure}
    \centering
    \includegraphics[width=1\linewidth]{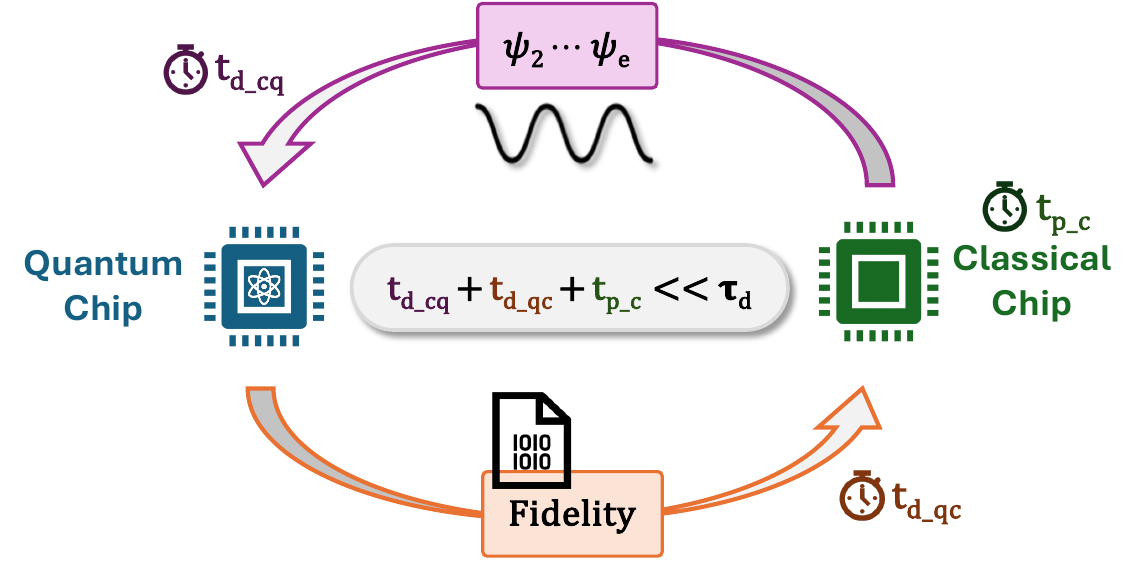}
    \caption{
\textbf{Quantum-Centric Supercomputing Architecture for Single-Copy State Reconstruction.}
This architecture enables non-destructive quantum state estimation by tightly integrating quantum and classical processors. An unknown quantum state is compared against a generated guess using a SWAP test. Fidelity feedback is processed classically to refine the guess, all within the qubit coherence time $\tau_d$. Key time components include: 
\textbf{$t_{d\_cq}$} – Time to send signals for dynamic circuit initialization from classical controller to quantum hardware; 
\textbf{$t_{d\_qc}$} – Time to retrieve measurement results (fidelity) from quantum to classical; 
\textbf{$t_{p\_c}$} – Time for classical model update and optimization. 
The total feedback loop time $t_{d\_cq} + t_{d\_qc} + t_{p\_c}$ must be significantly less than $\tau_d$ to preserve the original quantum state throughout iterative refinement.
}
    \label{fig:Q_C_arch}
\end{figure}

\subsection{Contribution}

In this work, we introduce a novel method for capturing \textit{quantum snapshots i.e. estimation of a quantum state at specific points during circuit execution}, and storing them in classical memory. Our approach is designed to be compatible with both current quantum hardware and future quantum-centric supercomputing architectures~\cite{IBMQuantumCentric2024}. It leverages dynamic circuits~\cite{IBMQuantumFeedforward2025} equipped with mid-circuit measurements~\cite{IBMRepeatUntilSuccess2025, PennyLaneMCM2025}, enabling quantum state reconstruction during runtime.
A key feature of our method is that it achieves this reconstruction without collapsing or destroying the quantum state from a single copy of the quantum state in the form of a quantum circuit. This makes it possible to access and store multiple quantum states at different points within a circuit, preserving coherence and enabling richer forms of intermediate analysis. The overall concept is illustrated in Figure~\ref{fig:quantum_bank}. These deposited states can later be retrieved or re-prepared on a quantum device by downstream applications.
By reconstructing and storing quantum states as classical vectors, our method lays the foundation for a form of classical quantum memory. These reconstructed states can be reloaded by quantum applications requiring access to previously encountered states, such as those employing standard Quantum RAM (QRAM) \cite{giovannetti_quantum_2008, phalak_quantum_2023} retrieval algorithms, thus enabling non-destructive reuse without relying on persistent physical quantum memory.
The primary contributions of this work are as follows:
%
    \circled{1}
    \textbf{Non-destructive Quantum Snapshots from a Single Copy at multiple intersections:} 
    We propose a novel methodology for capturing multiple simultaneous quantum state ``snapshots'' mid-circuit without destroying the original quantum system. This is accomplished by leveraging dynamic circuits equipped with mid-circuit measurements and controlled feedback, allowing quantum states to be reconstructed non-destructively from a single copy, thus, laying the foundation for real-time, introspective quantum computation.
\circled{2}
    \textbf{Quantum-centric Supercomputing Architecture:} 
    To support the proposed snapshot mechanism, we outline the capability requirement of a quantum-centric supercomputing architecture in Fig. \ref{fig:Q_C_arch}. It tightly couples quantum and classical systems, enabling rapid classical optimization in tandem with quantum evaluations, thereby mitigating decoherence limitations and supporting dynamic circuit execution. It enables multiple simultaneous snapshots at different circuit intersection as shown in Fig.\ref{fig:quantum_bank} within a single quantum execution, facilitating richer introspection and memory capabilities. 
\circled{3}
    \textbf{Proof of Concept on Real Quantum Hardware:} 
    We implement and validate a core component of our approach i.e. state reconstruction from fidelity measurements, on currently available IBM superconducting qubit hardware. We demonstrate unknown state reconstruction achieving fidelities exceeding 99.9\% in simulation and up to 99\% on hardware 
\circled{4}
    \textbf{Learning Quantum States Using Fidelity as the Only Signal:} 
    We develop and evaluate two complementary machine learning strategies: \textbf{(i)} a gradient-based deep neural network, and \textbf{(ii)} a gradient-free evolutionary strategy (QESwap), to reconstruct unknown quantum states using only fidelity (measured via the SWAP test) as feedback. Both methods are designed to optimize the state estimate purely from measurement statistics, without needing prior knowledge of the state or its density matrix.
\subsection{Viability of Quantum Snapshot with Dynamic Circuit (QSDC)}
\subsubsection{Work in Progress}
In our experiments, we show that Fidelity can be used as a metric for the successful reconstruction of unknown states. To extend the idea for dynamic circuits using mid-circuit measurement, we are actively developing a simulation-based implementation of the QSDC framework using PennyLane’s Catalyst \cite{noauthor_introducing_nodate}, a Just-In-Time compiler designed to support hybrid quantum-classical workflows. This direction is necessitated by the current limitations in quantum software development kits (SDKs), which lack mature support for high-level classical control required for dynamic quantum circuit execution. As a result, implementing iterative optimization procedures, such as those driven by deep neural networks or evolutionary strategies, remains non-trivial in current practical quantum hardware workflows.

\subsubsection{Hardware Limitations}
Additionally, hardware constraints such as limited qubit coherence times for technologies such as superconducting qubits for taking multiple simultaneous snapshots, restricted mid-circuit control and the absence of robust dynamic runtime environments further impedes the deployment of our method on existing quantum platforms. Despite these limitations, we are optimistic about future advancements. As platforms like Qiskit begin to support more expressive dynamic circuit primitives and low-latency classical feedback~\cite{noauthor_classical_nodate}, we plan to extend QSDC from simulation to hardware, enabling real-time, non-destructive quantum state reconstruction and storage.

\subsection{Paper Structure}
Section II discusses the theoretical background of dynamic circuits and mid-circuit measurements. We present the proposed QSDC in Section III showing the viability of our approach. Section IV describes the results and analysis and Section V concludes the paper.
\section{Background}

\subsection{Quantum Dynamic Circuits}

\begin{figure}
    \centering
    \includegraphics[width=0.5\linewidth]{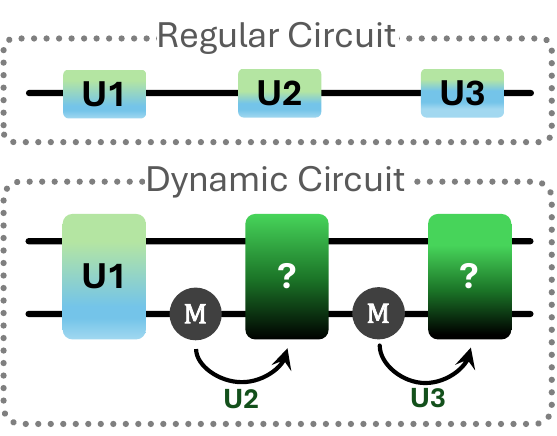}
    \caption{\textbf{Dynamic Quantum Circuit with mid-circuit measurement}
}
    \label{fig:dynamic_circuit}
\end{figure}

Quantum dynamic circuits \cite{noauthor_classical_nodate} enhance traditional circuits by enabling classical control flow during execution. Unlike static circuits, they support \textit{classical feedforward} \cite{noauthor_classical_nodate}, allowing quantum operations to adapt in real time based on measurement outcomes. This enables advanced techniques like error correction, teleportation, and adaptive algorithms. Dynamic circuits interleave quantum gates, mid-circuit measurements, and classical processing within the hardware's coherence time, enabling conditional branching and greater expressiveness.
\vspace{-0.1cm}
\subsection{Mid-Circuit Measurement}
Mid-circuit measurement\cite{wierichs_how_2024} involves measuring selected qubits during the execution of a quantum circuit, rather than only at the end. These measurements partially collapse the quantum state and produce classical information, which can be used to influence subsequent operations (Fig. \ref{fig:dynamic_circuit}). When paired with classical control logic, mid-circuit measurements enable adaptive quantum algorithms and efficient reuse of circuit components. To be effective, mid-circuit measurements must be low-latency and high-fidelity, minimizing their impact on the remaining quantum state. They are essential in enabling features such as conditional resets, real-time optimization, and quantum state learning.
Together, dynamic circuits and mid-circuit measurements form the architectural foundation for interactive quantum-classical computation. In this work, we leverage this capability to learn unknown quantum states through adaptive circuit design and measurement-based feedback. 
\vspace{-.3cm}
\section{Methods}\label{methods}
 
\subsection{Quantum Snapshot with Dynamic Circuit (QSDC)}
We present our framework, QSDC, for identifying an unknown quantum state at any intermediate stage of a quantum circuit. Once estimated, the state is stored classically in conventional memory for future reconstruction.
QSDC operates via hybrid quantum-classical interaction, assuming a classical quantum controller co-located with the quantum processor to enable fast classical decision-making within the coherence time of the qubits. The unknown quantum state is represented as a quantum circuit—consistent with modern quantum computing paradigms. For demonstration, we generate a random quantum state using the \texttt{StatePrep} primitive in PennyLane (Fig. ~\ref{fig:process}A). A candidate state is then initialized by our classical ML model (Fig. ~\ref{fig:process}B) and loaded into the circuit using \texttt{StatePrep}, accompanied by an ancilla qubit to perform the SWAP test.
Assuming sufficiently long coherence times (Fig. ~\ref{fig:Q_C_arch}), iterative classical computations can be interleaved with quantum operations, enabling repeated state estimation within a single quantum circuit. While current qubit technologies (e.g., superconducting and Trapped-Ion qubits) lack the coherence time and classical integration required for such real-time feedback, we are simulating this architecture under the outlined assumptions as part of ongoing work.
In the QSDC procedure, the generated state $\ket{\psi}$ is iteratively refined using gradient-free or gradient-based optimization, based on fidelity feedback from the SWAP test. After each evaluation, qubits $R_1$ to $R_N$ are reset and reinitialized with the updated candidate state. The ancilla is also reset, and SWAP gates are reapplied between the unknown state ($T_1$ to $T_N$) and the candidate state ($R_1$ to $R_N$) to reassess fidelity. This process repeats for a fixed number of epochs. The method assumes qubit state preservation throughout hybrid optimization—an aspirational but increasingly feasible condition as coherence times approach the millisecond regime, potentially enabling accurate reconstruction from a single copy on real hardware.
We employ two estimation strategies for quantum state reconstruction:  
(a)~a \textbf{gradient-based} approach using a deep learning generator trained via backpropagation, and  
(b)~a \textbf{gradient-free} approach utilizing evolutionary strategies for optimization.
These strategies are applied across three different quantum state representations:  
(i)~\textbf{State vector}—used for pure states,  
(ii)~\textbf{Density matrix}—which generalizes to both pure and mixed states, and  
(iii)~\textbf{Unitary matrix}—describing the transformation from the initial state $\ket{0}^{\otimes n}$ to the target state $\ket{\psi}$.
For the \textit{state vector}, we estimate $\ket{\psi} \in \mathbb{C}^{2^n}$ directly. In the case of \textit{density} and \textit{unitary} matrices, we reconstruct $2^n \times 2^n$ complex-valued matrices, subject to necessary physical constraints—Hermiticity and unit trace for density matrices, and unitarity for unitaries.
We begin by presenting the gradient-based estimation method for the state vector, followed by its gradient-free counterpart. This methodology is then extended to density and unitary matrix reconstruction, with appropriate modifications to accommodate their structural and dimensional complexities. The procedure for state vector estimation forms the foundation for these extensions.

\subsection{Gradient Based Method: Neural Network as Generator}

\subsubsection{\textbf{Problem Formulation}}

The objective of this method is to reconstruct an unknown quantum state vector $\ket{\psi} \in \mathbb{C}^{2^n}$ without prior knowledge of its structure, using a classical deep neural network. The goal is to train the model to generate a state $\ket{\phi}$ such that the fidelity
\(F = |\bra{\psi}\ket{\phi_u}|^2\)
approaches unity—implying that the generated state is indistinguishable from the target up to a global phase \cite{Nielsen_Chuang_2010}.
To achieve this, we design a deep neural network generator that learns to produce quantum states matching the unknown target. The SWAP test serves as a fidelity-based loss function, providing feedback on the similarity between $\ket{\phi}$ and $\ket{\psi}$. This feedback iteratively guides the generator toward producing a state that maximizes fidelity, ideally reaching $F \approx 1$.
The key components of the QSDC are as follows.
\subsubsection{\textbf{Unknown Quantum State Generation}}\label{unknown_state_prep}

To simulate an arbitrary, unknown quantum state $\ket{\phi_u}$ over $n$ qubits, we sample a normalized random state vector from the complex unit sphere in $\mathbb{C}^{2^n}$. The real and imaginary components of each amplitude are independently drawn from a standard normal distribution and normalized to unit length, i.e., $\ket{\phi_u} = (\vec{r} + i \vec{i}) / \|\vec{r} + i \vec{i}\|$, where $\vec{r}, \vec{i} \in \mathbb{R}^{2^n}$ and $\vec{r}, \vec{i} \sim \mathcal{N}(0,1)$. This state acts as a proxy for unknown quantum states encountered in practical quantum computations. Its corresponding quantum circuit is synthesized using the Mottonen state preparation method~\cite{mottonen_transformation_2004}, allowing efficient loading of arbitrary quantum states for reconstruction within the QSDC framework.

\subsubsection{\textbf{Generator Architecture}}\label{generator_arch}

The quantum state generator is implemented as a deep, fully connected neural network using PyTorch. It takes a 256-dimensional latent vector $\mathbf{z}$ as input and outputs a real-valued vector of dimension $2 \times \texttt{dim}$, where $\texttt{dim}$ is the number of complex amplitudes in the target state. This output is interpreted as the concatenation of the real and imaginary parts of the quantum state and is subsequently normalized. The architecture consists of six linear layers with GELU activations (except at the output layer). The trained network outputs a normalized candidate state vector $\ket{\psi}$, which is then encoded into a quantum circuit using the Mottonen state preparation method~\cite{mottonen_transformation_2004}.

\subsubsection{\textbf{Swap Test for Fidelity Estimation}}\label{swap_test}

To evaluate the similarity between the generated state $\ket{\psi}$ and the unknown target $\ket{\phi_u}$, we estimate fidelity using a SWAP test, implemented via dynamic circuits with mid-circuit measurements to preserve the unknown state during evaluation. The fidelity is defined as $F(\psi, \phi_u) = |\langle \psi | \phi_u \rangle|^2$. In the SWAP test, let $P(0)$ be the probability of measuring the ancilla qubit in state $\ket{0}$; then $F = 2P(0) - 1 = \langle Z \rangle$, where $\langle Z \rangle$ is the expectation value of the Pauli-$Z$ operator on the ancilla qubit.

The SWAP test circuit uses $2n+1$ qubits: (1) Qubits $1$ to $n$ are initialized with $\ket{\phi_u}$ once; (2) Qubits $n+1$ to $2n$ are reinitialized with $\ket{\psi}$ at each iteration; (3) Qubit $0$ serves as the ancilla for controlled-SWAP operations. After each fidelity evaluation, qubits $n+1$ to $2n$ and the ancilla are reset and updated with the latest candidate state. The fidelity $\langle Z \rangle = |\langle \psi | \phi_u \rangle|^2$ is used as the training objective.

Due to the SWAP test's non-differentiable nature, we implement a custom autograd function using finite-difference approximation. In the forward pass, the real and imaginary parts of the generated vector form a normalized complex state; fidelity is computed against $\ket{\phi_u}$. During the backward pass, symmetric finite differences estimate gradients, enabling training via standard gradient-based optimization.

\subsubsection{\textbf{Training Procedure}}\label{training_procedure}

The fidelity score between the generated state $\ket{\psi}$ and the target state $\ket{\phi_u}$ defines the loss function $\mathcal{L} = 1 - F$, where $F = |\langle \psi | \phi_u \rangle|^2$. This loss guides the generator toward maximizing overlap with $\ket{\phi_u}$. Due to the non-differentiability of quantum operations like the SWAP test, gradients are estimated via finite-difference approximation using a custom autograd function.

Training is performed using the Adam optimizer with a learning rate of $1 \times 10^{-4}$. At each step, a latent vector $\mathbf{z} \sim \mathcal{U}(0, 1)^{256}$ is sampled and mapped to a normalized complex quantum state by the generator. Fidelity is evaluated, and the loss is backpropagated using finite-difference gradients. To improve stability, gradients are scaled by a constant \texttt{scaling\_factor} before optimizer updates. Training proceeds for a fixed number of epochs (depending on the number of qubits), and the highest fidelity achieved is recorded. After training, the final state is classically compared to the target via $|\langle \psi | \phi_u \rangle|^2$, with $F \approx 1$ indicating successful reconstruction.

\subsection{Gradient Free Method: Evolutionary Strategy}\label{grad_free}

Gradient-based methods achieved high fidelities in noiseless simulations but required extensive hyperparameter tuning and were unreliable in noisy settings due to unstable gradients, leading to slow convergence. For higher-dimensional states, fidelity often plateaued near 0.85—below the target of 0.99. A key challenge was gradient estimation, as the non-differentiable SWAP test necessitated costly finite-difference methods. To address this, we explored gradient-free approaches, finding Evolutionary Strategies (ES) \cite{salimans_evolution_2017} to outperform gradient-based methods in both fidelity and convergence. We introduce \textbf{QESwap (Quantum Evolutionary Strategy with SWAP Fidelity)}, a derivative-free framework for quantum state reconstruction using SWAP-test fidelity. By leveraging Gaussian perturbations and population-based search, QESwap enables non-destructive, single-copy learning, making it ideal for mid-circuit state reconstruction and dynamic quantum memory. To our knowledge, it is the first ES approach to incorporate SWAP fidelity in a hybrid quantum-classical loop.
Our method is implemented using \texttt{PennyLane} and \texttt{PyTorch}, combining quantum and classical components to optimize representations of quantum states—whether as state vectors, density matrices, or unitaries. Without loss of generality, we describe the approach below using the state vector representation.

\subsubsection{\textbf{Problem Formulation}} Let the unknown quantum state be denoted by \( |\phi_{\text{unknown}}\rangle \), a normalized complex vector in a \( 2^n \)-dimensional Hilbert space for \( n \) qubits. The objective is to estimate a parametrized state \( |\psi_{\text{gen}}(\mathbf{w})\rangle \) that maximizes the fidelity
\(
F = \left| \langle \phi_{\text{u}} \mid \psi_{\text{gen}}(\mathbf{w}) \rangle \right|^2.
\)

\subsubsection{\textbf{Evolutionary Strategy Optimization}} The Evolutionary Strategy (ES) algorithm searches for optimal parameters \( \mathbf{w} \in \mathbb{R}^{2 \cdot 2^n} \), representing the flattened real and imaginary components of a candidate quantum state. No assumptions are made about the structure or sparsity of the target state.
At each iteration, a population of \( N = 50 \) perturbed candidates is generated by adding Gaussian noise to \( \mathbf{w} \). Each candidate is interpreted as a complex vector, normalized to unit norm to ensure physical validity, and evaluated using the SWAP test to compute fidelity.
The parameters are updated using:
\(
\mathbf{w} \leftarrow \mathbf{w} + \frac{\alpha}{N \sigma} \sum_i A_i \cdot \mathbf{z}_i,
\)
where \( \alpha = 0.05 \) is the learning rate, \( \sigma = 0.1 \) is the noise scale, \( A_i \) is the standardized advantage, and \( \mathbf{z}_i \) is the noise vector for the \( i \)-th candidate.
Training terminates upon reaching a fidelity threshold (e.g., 0.95 or 0.99) or a maximum of 100 iterations. All candidate states are normalized prior to fidelity evaluation.

\subsubsection{\textbf{Experimental Protocol}} We evaluate QESwap on 1 to 6-qubit systems using 1000 trials per configuration with randomly sampled target states. For each trial, fidelity and convergence epochs are recorded for thresholds \( F \geq 0.95 \) and \( F \geq 0.99 \). Experiments are run on both noisy and noiseless simulators, with hardware results limited to the single-qubit case (Table~\ref{tab:all_results}).

\subsection{Computing Hardware to enable QSDC}
To enable QSDC and single-copy, non-destructive quantum snapshots on real hardware, we propose a quantum-centric architecture with tight quantum-classical feedback (Fig.~\ref{fig:Q_C_arch}). A low-latency classical controller—e.g., an ASIC—co-located with the quantum processor updates the state generation model after each SWAP test and reinitializes the circuit within the qubit coherence window. This requires mid-circuit reset, preparation, and measurement, now supported by platforms like IBM’s dynamic circuit SDKs~\cite{IBMQuantumCentric2024}. Our experimental results demonstrate that the reconstruction process typically converges within a few optimization steps (e.g., 3 epochs for single-qubit states), suggesting that an entire snapshot can be captured within a few milliseconds—well within the coherence times of many quantum hardware platforms such as neutral atoms and trapped ions. While current superconducting qubits may still pose timing challenges, anticipated improvements could make this feasible in the near future. This rapid convergence mitigates decoherence limitations and makes single-copy estimation practically viable.
\subsection{Storage and Retrieval of Quantum Snapshots:} Once high fidelity is achieved, the candidate state $\ket{\psi}$ is stored classically (Fig. \ref{fig:quantum_bank}) and can be re-prepared on quantum hardware using standard circuit synthesis techniques compatible with QRAM architectures.



\section{Results and Analysis}
We validated our approach by reconstructing unknown quantum states using SWAP test fidelity-based optimization without mid-circuit measurements, reinitializing the circuit at each step. The approximate QSDC method was tested on the noisy 127-qubit \texttt{ibm\_sherbrooke} and in noiseless simulations using AerSimulator in PennyLane.

\subsection{Real Hardware Results}
We validated QESwap on known single-qubit states, including \( \ket{0} = [1, 0] \), \( \ket{1} = [0, 1] \), and the Hadamard state \( \frac{1}{\sqrt{2}}[1, 1] \), achieving fidelity 1 within 3 epochs. Due to hardware constraints, large-scale trials with 100 random states were limited to noisy and noiseless simulations, discussed in the following sections.


\begin{table*}[!htbp]
\centering
\fontsize{7.0pt}{9.5pt}\selectfont
\caption{Average epochs to reach fidelity $> 0.99$ for 1, 2, and 3-qubit circuits across simulated and real backends computed over 100 random circuits and shown only for cases that successfully exceeded the threshold.}
\label{tab:table_master_result}
\centering
\begin{tabular}{c|c|cccccccccccc|cc}
\hline
\multirow{3}{*}{\textbf{\begin{tabular}[c]{@{}c@{}}Estimation\\ Strategies\end{tabular}}}                   & \multirow{3}{*}{\textbf{\begin{tabular}[c]{@{}c@{}}Regenerating\\ Methods\end{tabular}}} & \multicolumn{6}{c|}{\textbf{Noiseless Simulations}}                                                                                                          & \multicolumn{6}{c|}{\textbf{Noisy Simulations}}                                                                                                  & \multicolumn{2}{c}{\textbf{Real Hardware}} \\ \cline{3-14}
                                                                                                            &                                                                                          & \multicolumn{2}{c|}{\textbf{1 qubit}}              & \multicolumn{2}{c|}{\textbf{2 qubits}}             & \multicolumn{2}{c|}{\textbf{3 qubits}}             & \multicolumn{2}{c|}{\textbf{1 qubit}}              & \multicolumn{2}{c|}{\textbf{2 qubits}}             & \multicolumn{2}{c|}{\textbf{3 qubits}} & \multicolumn{2}{c}{\textbf{1 qubit}}       \\ \cline{3-16} 
                                                                                                            &                                                                                          & \textbf{\# E} & \multicolumn{1}{c|}{\textbf{Fid.}} & \textbf{\# E} & \multicolumn{1}{c|}{\textbf{Fid.}} & \textbf{\# E} & \multicolumn{1}{c|}{\textbf{Fid.}} & \textbf{\# E} & \multicolumn{1}{c|}{\textbf{Fid.}} & \textbf{\# E} & \multicolumn{1}{c|}{\textbf{Fid.}} & \textbf{\# E}      & \textbf{Fid.}     & \textbf{\# E}        & \textbf{Fid.}       \\ \hline
\multirow{3}{*}{\textbf{\begin{tabular}[c]{@{}c@{}}Gradient Based \\ (Deep Neural\\ Network)\end{tabular}}} & \textbf{State Vector}                                                                    & 5             & \multicolumn{1}{c|}{0.9999}        & 12            & \multicolumn{1}{c|}{0.9997}        & 17            & \multicolumn{1}{c|}{0.9997}        & 27            & \multicolumn{1}{c|}{0.9927}        & \multicolumn{4}{c|}{NA}                                                                     & \multicolumn{2}{c}{\multirow{3}{*}{NA}}    \\ \cline{2-14}
                                                                                                            & \textbf{Unitary Matrix}                                                                  & 32            & \multicolumn{1}{c|}{0.9999}        & 48            & \multicolumn{1}{c|}{0.9999}        & 75            & \multicolumn{1}{c|}{0.9999}        & \multicolumn{6}{c|}{NA}                                                                                                                          & \multicolumn{2}{c}{}                       \\ \cline{2-14}
                                                                                                            & \textbf{Density Matrix}                                                                  & \multicolumn{12}{c|}{NA}                                                                                                                                                                                                                                                                                        & \multicolumn{2}{c}{}                       \\ \hline
\multirow{3}{*}{\textbf{\begin{tabular}[c]{@{}c@{}}Gradient Free\\ (Evolutionary\\ Strategy)\end{tabular}}} & \textbf{State Vector}                                                                    & 5             & \multicolumn{1}{c|}{0.9955}        & 7             & \multicolumn{1}{c|}{0.9951}        & 13            & \multicolumn{1}{c|}{0.9928}        & 5             & \multicolumn{1}{c|}{0.9963}        & 8             & \multicolumn{1}{c|}{0.9945}        & 26                 & 0.9906            & 3                    & 0.99                \\ \cline{2-16} 
                                                                                                            & \textbf{Unitary Matrix}                                                                  & 8             & \multicolumn{1}{c|}{0.9978}        & 11            & \multicolumn{1}{c|}{0.9949}        & 19            & \multicolumn{1}{c|}{0.9937}        & 59            & \multicolumn{1}{c|}{0.99}          & \multicolumn{4}{c|}{NA}                                                                     & \multicolumn{2}{c}{\multirow{2}{*}{NA}}    \\ \cline{2-14}
                                                                                                            & \textbf{Density Matrix}                                                                  & \multicolumn{12}{c|}{NA}                                                                                                                                                                                                                                                                                        & \multicolumn{2}{c}{}                       \\ \hline
\end{tabular}
\label{tab:all_results}
\end{table*}

\subsection{Unknown State Preparation with High-Quality Entanglement and Faithful Reconstruction }
We prepared 100 random unknown states \( \ket{\phi_u} \) as described in Section~\ref{unknown_state_prep} and used them across all experiments unless noted otherwise. To quantify entanglement, we computed the von Neumann entropy via bipartition: \( \rho_A = \mathrm{Tr}_B(\lvert\psi\rangle\langle\psi\rvert) \), with 
\(
S(\rho_A) = -\mathrm{Tr}(\rho_A \log \rho_A),
\)
capturing entanglement across subsystems. The dataset included both highly and weakly entangled (matrix product) states. Reconstructed states showed entanglement entropy closely matching the targets, indicating robustness. On comparing the entanglement entropy across qubit counts and circuit index, we observe a higher entropy for 8-qubit circuits for the target states. On sorting the circuits by the index size, we find a close alignment with the generated states across different system sizes.


We perform experiments on 100 randomly sampled states and analyze the fidelity. In noiseless simulations, both neural network (NN) and evolutionary strategy (ES) approaches achieved fidelities near 1. In noisy settings, performance is degraded due to gradient instability, particularly for NN. To manage computational cost, NN was tested up to 2 qubits and ES up to 3. ES demonstrated faster convergence and was easier to tune (Table~\ref{tab:all_results}).
For unitary matrix reconstruction, fidelity remained high in noiseless simulations but degraded rapidly under noise as qubit count increased (Table~\ref{tab:table_master_result}), limiting further noisy evaluations. In density matrix reconstruction using NN, fidelity plateaued at ~0.8 even in noiseless 2-qubit cases despite tuning efforts. This was traced to the SWAP test’s inability to fully capture overlap for mixed states, making it unsuitable for mixed-state fidelity estimation.
To confirm our optimization method was not the bottleneck, we evaluated using Uhlmann fidelity—which accurately captures overlap for mixed states but requires access to the target density matrix. Under this metric, our method achieved near-perfect fidelity, confirming its effectiveness. We conclude that our approach is well-suited for pure state reconstruction, while SWAP-test-based fidelity limits its applicability to mixed states.

While our approach excels at reconstructing pure quantum states using SWAP-test-based fidelity optimization, its extension to mixed states reveals fundamental limitations. This stems from the SWAP test estimating the Hilbert-Schmidt inner product, $\operatorname{Tr}(\rho \sigma)$, rather than true quantum fidelity---leading to overestimated similarity and limited operational relevance for tasks like discrimination or verification. Consequently, models trained on this metric may converge to states that appear similar mathematically but are physically dissimilar in trace distance or Uhlmann fidelity. Furthermore, the SWAP test has limited discrimination power for mixed states, with a success probability bounded by $P_{\text{success}} \leq \frac{3}{4}$, while optimal Helstrom measurements can achieve $P_{\text{success}} = 1 - \frac{1}{2} \operatorname{Tr}|\rho - \sigma|$. These gaps underscore a structural and informational bottleneck in extending our method to general mixed states. Nonetheless, this limitation is largely theoretical in the context of near-term quantum computing, where most target states---such as Bell or GHZ states---are pure by design. Even with noise-induced mixing, our method’s high-fidelity pure-state reconstructions serve as effective approximations for debugging, introspection, and quantum memory. Use cases like QRAM and modular circuits inherently assume pure inputs and outputs, aligning well with our framework’s strengths. As such, the pure-state focus is not a practical constraint but a reflection of prevailing algorithmic assumptions in quantum computing.

\textbf{Training Performance Analysis and Comparison Between Learning Strategies:}
To assess convergence performance, we measured the number of epochs required to achieve fidelity \( > 0.99 \) across various qubit configurations and environments. Table~\ref{tab:all_results} summarizes average epoch counts for different estimation strategies under noiseless, noisy, and hardware settings.
In noiseless simulations, the gradient-based neural network (NN) reached \( > 0.999 \) fidelity within 5–17 epochs for state vectors, and 32–75 epochs for unitaries. The gradient-free evolutionary strategy (ES) converged faster, requiring only 5–13 epochs for state vectors and fewer than 20 for unitaries.
Under noise, NN models showed slower, more variable convergence, while ES remained robust, reaching \( > 0.99 \) fidelity in under 30 epochs for up to 3-qubit states. On real hardware, ES successfully reconstructed single-qubit states in just 3 epochs, highlighting its efficiency and practicality.
Overall, ES consistently demonstrated faster and more stable convergence, particularly in noisy and resource-constrained settings.

\vspace{-.3cm}

\section{Conclusion}
We presented QSDC, a hybrid quantum-classical framework for non-destructive, single-copy quantum state reconstruction using SWAP-test-based fidelity. By combining dynamic circuits with learning-based optimization, QSDC enables multiple mid-circuit snapshots within a single copy of quantum circuit and there after classical storage of those quantum states.
Simulations and hardware trials validate its effectiveness, with evolutionary strategies showing robust performance in noisy settings. While limited to pure states, QSDC lays the groundwork for introspective, memory-enabled quantum computing in future dynamic hardware environments.

\section*{Acknowledgment}
The work is supported in parts by the National Science Foundation (NSF) (CNS-2129675 and CCF-2210963) and gifts from Intel.

\bibliography{ref_zotero_qram,ref}
\bibliographystyle{ieeetr}


\end{document}